\documentclass[aps,twocolumn,groupedaddress,amsmath,amssymb]{revtex4}
\usepackage{amsmath, dsfont}
\usepackage{amssymb}
\usepackage[dvips]{graphics}
\usepackage{epsfig}
\usepackage{calc}
\usepackage{amsfonts}
\usepackage{graphicx}
\usepackage[ansinew]{inputenc}

\begin{document}
\setcounter{page}{1}
\title[]{Topological black holes for Einstein-Gauss-Bonnet gravity  with a nonminimal scalar field}

\author{Mois\'es Bravo Gaete}\email{mbravog-at-inst-mat.utalca.cl}
\affiliation{Instituto de Matem\'atica y F\'{\i}sica, Universidad de
Talca, Casilla 747, Talca, Chile.}

\author{Mokhtar Hassa\"{\i}ne}\email{hassaine-at-inst-mat.utalca.cl}
\affiliation{Instituto de Matem\'atica y F\'{\i}sica, Universidad de
Talca, Casilla 747, Talca, Chile.}

\begin{abstract}
We consider the Einstein-Gauss-Bonnet gravity with a negative cosmological constant together with a source given by a
scalar field nonminimally coupled in arbitrary dimension $D$. For a certain election of
the cosmological and Gauss-Bonnet coupling constants, we derive two classes of AdS black hole solutions whose horizon is
planar. The first family of black holes obtained for a particular value of the nonminimal
coupling parameter only depends on a constant $M$, and the scalar field vanishes as $M=0$.
The second class of solutions corresponds to a two-parametric (with constants $M$ and $A$) black hole stealth configuration, that is a nontrivial scalar
field with a black hole metric such that both side (gravity and matter parts) of the Einstein equations vanishes. In this case, in the vanishing $M$,
the solution reduces to a stealth scalar field on the pure AdS metric. We note that the existence
of these two classes of solutions is inherent of the particular choice of the coupling constants and, they can not
be promoted to spherical or hyperboloid black hole solutions in a standard fashion. In the last part,
we add to the original action some exacts $(D-1)-$forms coupled to the scalar field. The direct benefit of introducing such extra fields
is to obtain black hole solutions with planar horizon for arbitrary value of the nonminimal coupling parameter.

\end{abstract}

\maketitle

\section{Introduction}
The anti-de Sitter/Conformal Field Theory (AdS/CFT) correspondence is a power tool that permits to analyze
strongly coupled systems by mapping them into some higher-dimensional gravity theories and establishing a sort of dictionary between both
theories \cite{Maldacena:1997re}. Recently, these ideas have been extended to non-relativistic physics in order to describe certain condensed
matter systems (e.g.~see \cite{Kovchegov:2011ub,Hartnoll:2009sz} for recent reviews) or in order to gain a better understanding of some unconventional superconductors \cite{Horowitz:2010gk,Hartnoll:2008kx,Herzog:2009xv}. In this last case, the minimal ingredients in the gravity side are given by the Einstein-Hilbert action with a negative cosmological constant together with a charged self-interacting (complex) scalar field with the Maxwell term. Moreover, in order to reproduce the superconductor phase diagram, the system must admit black holes with scalar
hair at low temperature and this hair must disappear at high temperature. However, this problem of finding black hole solutions with scalar field is rendered difficult by the various no-hair theorems existing in the current literature, see e.g.~\cite{Bekenstein:1998aw}. Nevertheless, such no-go theorems can be avoided considering scalar fields nonminimally coupled to gravity \cite{Bekenstein:1974sf, Bocharova:1970}. We will precisely consider this
kind of matter source in order to escape the traditional no-hair theorems. For the gravity Lagrangian, we will be concerned with the  Einstein-Gauss-Bonnet action with a negative cosmological constant. This choice is motivated by the recent interest on holographic superconductors in
Einstein-Gauss-Bonnet gravity. Indeed, holographic superconductors in  such gravity theory has been studied intensively with the purpose of
analyzing the effects of the Gauss-Bonnet coupling constant on the critical temperature and on the condensate, see e.g. \cite{Barclay:2010up} and \cite{Yao:2013sha, Cui:2013uha}.

More specifically, we will consider a matter action given by a particular Einstein-Gauss-Bonnet gravity action
together with a self-interacting scalar field nonminimally coupled to gravity, and look for black holes; for good reviews on Einstein-Gauss-Bonnet black holes, see e.g. \cite{Charmousis:2008kc} and \cite{Garraffo:2008hu}. For this model, we derive two classes
of black hole solutions with planar base manifold and only for particular values of the nonminimal coupling parameter. Note that the first examples of topological black hole in GR 
were discussed in \cite{Lemos}. In our case, for $\xi=(D-2)/(4D)$ and $\xi=(D-1)/(4D+4)$,
we obtain AdS
black holes whose metrics resemble the Schwarzschild-AdS-Tangherlini spacetime. In both case and in contrast with the BBMB solution \cite{Bekenstein:1974sf, Bocharova:1970}, the scalar field does not diverge at the horizon. This is due to the presence of the negative cosmological constant as it occurs for the known black hole solutions with scalar fields in four dimensions  \cite{Martinez:2002ru, Martinez:2005di, Anabalon:2012tu}. For the first family of solutions, the scalar field depends on the mass constant $M$ and vanishes identically as $M=0$ yielding to the purely AdS solution without source. The second class of solution is at our knowledge the first example of
higher-dimensional black hole stealth configuration. We mean a nontrivial scalar field together with a black hole metric such that both side of the Einstein equations (gravity and matter parts) vanishes. Up to now, the only known black hole stealth was derived in three dimensions for a conformal scalar field \cite{AyonBeato:2004ig} on the BTZ black hole \cite{Banados:1992wn}. Finally, in the last part, we add to our original action  $(D-2)$ dynamical fields which are exact $(D-1)$-forms coupled with the scalar field in order to relax the restriction on the nonminimal coupling parameter. In this case, we will establish the existence of black hole solutions for arbitrary $\xi$ reducing to the pure scalar field solutions as $\xi=(D-2)/(4D)$ or $\xi=(D-1)/(4D+4)$. There also exist a value of the nonminimal parameter $\xi=1/8$ giving rise to a pure axionic solution, that is a solution where the contribution of the  scalar fields disappears.

The plan of the paper is organized as follows. In the next section, we will present the model and the associated field equations, and
derive two classes of solutions. In Sec. III, we will add to our starting action, some axionic fields coupled to the scalar field and obtain black hole solutions generalizing in some sense those obtained in the previous section. Finally, the last section is devoted to the conclusions and the further works.

\section{Topological black holes for Einstein-Gauss-Bonnet gravity with a scalar field nonminimally coupled}
 We consider the following action in arbitrary $D$ dimension with $D\geq 5$,
\begin{align}\label{action1}
S=&\int d^{D}x\,\sqrt{-g}\, \Big[\frac{1}2\left(R-2\Lambda+\alpha\,\mathcal{L}_{GB}\right)\Big]\nonumber\\
&-\int d^{D}x\,\sqrt{-g}\, \Big[\frac{1}2\partial_{\mu}\Phi\partial^{\mu}\Phi+\frac{\xi}2
R\Phi^2 +U(\Phi)\Big],
\end{align}
where $\mathcal{L}_{GB}$ corresponds to the Gauss-Bonnet Lagrangian
$$
\mathcal{L}_{GB}=R^{2}-4\,R_{\mu \nu}R^{\mu \nu}+R_{\alpha\beta\mu\nu}R^{\alpha\beta\mu\nu}.
$$
Here, we have normalized the Newton coupling constant $G$ as $8\pi G=1$, and set the AdS radius $l$ to unity $l=1$. The gravity part of (\ref{action1}) corresponds to
the Einstein-Gauss-Bonnet action with a
cosmological constant $\Lambda$, while the matter source is given by a self-interacting scalar field $\Phi$ nonminimally coupled to
the scalar curvature $R$ through the nonminimal coupling parameter $\xi$. The potential $U(\Phi)$  is given by a mass term
\begin{equation}\label{potstealth}
U(\Phi)=\frac{8\,\xi\,D\,(D-1)}{(1-4\xi)^{2}}\,(\xi-\xi_{D})(\xi-\xi_{D+1}){\Phi}^{2},
\end{equation}
where $\xi_{D}$ denotes the conformal coupling in $D$ dimensions
\begin{equation}\label{conformal}
\xi_{D}=\frac{D-2}{4(D-1)}.
\end{equation}
The choice of such potential will be justified in the discussion. We also note that for the conformal couplings in $D$ and $D+1$ dimensions, the potential vanishes identically.

The field equations obtained by varying the action with respect with the metric and the scalar field read
\begin{subequations}
\label{eqsmotion}
\begin{eqnarray}
&&G_{\mu\nu}+\Lambda g_{\mu\nu}+\alpha K_{\mu\nu}=T_{\mu\nu},\\
&&\Box \Phi = \xi R \Phi+\frac{d U}{d\Phi},
\end{eqnarray}
\end{subequations}
where the expression of the Gauss-Bonnet tensor $K_{\mu\nu}$ is
\begin{eqnarray}
K_{\mu\nu}=& 2\big(RR_{\mu\nu}-2R_{\mu\rho}R^{\rho}_{\,\,\nu}-2R^{\rho\sigma}R_{\mu\rho\nu\sigma}
+R_{\mu}^{\,\,\rho\sigma\gamma}R_{\nu\rho\sigma\gamma}\big)\nonumber\\
&-\frac{1}{2}\, g_{\mu\nu}\mathcal{L}_{GB},
\label{GBmunu}
\end{eqnarray}
and the stress tensor associated to the variation of the scalar field is given by
\begin{eqnarray}
\label{tmunusf}
T_{\mu \nu}=&\partial_{\mu}\Phi\partial_{\nu} \Phi-g_{\mu
\nu}\Big(\frac{1}{2}\,\partial_{\sigma}\Phi\partial^{\sigma}\Phi+U(\Phi)\Big)\nonumber\\
&+\xi\left(g_{\mu \nu}\Box-\nabla_{\mu}\nabla_{\nu}+G_{\mu\nu}\right)\Phi^{2}.
\end{eqnarray}

In what follows, we will fix the value of the cosmological constant and the Gauss-Bonnet coupling constant $\alpha$  in terms of the dimension $D$ as
\begin{eqnarray}
\label{L2}
&&\Lambda=-\frac{(D-1)(D-2)}{4},\; \alpha=\frac{1}{2(D-3)(D-4)}
\end{eqnarray}
As a consequence of this election, the gravity part of the action (\ref{action1}) becomes proportional to
\begin{eqnarray}
\int d^D x\,\sqrt{-g}\left(R+\frac{(D-1)(D-2)}{2}+\frac{{\cal L}_{GB}}{2(D-3)(D-4)}\right).
\label{actionBHscan}
\end{eqnarray}
In five dimensions, this action corresponds to a Chern-Simons action which is a particular case of the
Lovelock action. This latter can be viewed as a generalization of the Einstein
gravity in arbitrary dimension yielding at most to second-order field equations for the metric. We will come to this point
in the discussion when commenting that the solutions derived here in the Einstein-Gauss-Bonnet case can be
extended to a particular class of higher-order Lovelock gravity \cite{BH}.

\subsection{Topological black hole solutions}
As shown now for the particular choice of the coupling constants (\ref{L2}), or equivalently for a gravity action given by (\ref{actionBHscan}), we
will derive two classes of topological black hole solutions of Eqs. (\ref{eqsmotion}). In both cases, the metric solutions which have a planar base
manifold resemble to the topological Schwarzschild-AdS-Tangherlini spacetime.

For a value of the nonminimal coupling parameter given by
\begin{eqnarray}
\label{xi2}
\xi=\xi^{\tiny{\mbox{b.h}}}_{D}:=\frac{D-2}{4D},
\end{eqnarray}
which implies that the potential (\ref{potstealth}) becomes
\begin{eqnarray}
\label{upot2}
U(\Phi)=\frac{(D-2)^2}{32}\Phi^2,
\end{eqnarray}
a solution of the field equations (\ref{eqsmotion}) in this case is given by
\begin{eqnarray} \label{solaxfield2}
&&ds^{2}=-\Big(r^2-\frac{M}{r^{\frac{D-6}{2}}}\Big)dt^{2}+\frac{dr^2}{\Big(r^2-\frac{M}{r^{\frac{D-6}{2}}}\Big)}+{r^{2}}d\vec{x}_{D-2}^2
\nonumber\\
&& \Phi(r)=2\sqrt{\frac{DM}{D-2}}\,r^{\frac{2-D}{4}},
\end{eqnarray}
where $\vec{x}$ denotes a $(D-2)-$dimensional vector. Various comments can be made concerning this solution. Firstly, the scalar field is real provided that the constant $M$ is positive, and is
well-defined at the horizon while diverging at the singularity $r=0$. The solution resembles to the topological Schwarzschild-AdS solution
with planar horizon. We also stress from now that, at the vanishing mass limit $M=0$, the scalar field vanishes and the metric is nothing but the
AdS metric written in Poincar\'e coordinates satisfying the pure gravity equations $G_{\mu\nu}+\Lambda g_{\mu\nu}+\alpha K_{\mu\nu}=0$  at the point (\ref{L2}). It is also interesting to point out  that the spherical or hyperboloid versions of the metric do not accommodate such a source unless the constant $M=0$. However, in this case, the scalar field vanishes identically and the spacetime
geometry is nothing but the AdS metric solving trivially the gravity equation $G_{\mu\nu}+\Lambda g_{\mu\nu}+\alpha K_{\mu\nu}=0$ with a cosmological constant
and Gauss-Bonnet coupling given by (\ref{L2}).


\subsection{Topological black hole stealth solutions}

Interestingly enough, there exists another value of the nonminimal parameter that yields to an interesting solution, namely a {\it stealth configuration}. By
stealth configuration, we mean a nontrivial solution (that is  a solution with a non-constant and non-vanishing scalar field) of the stealth equations
\begin{equation}
G_{\mu\nu}+\Lambda g_{\mu\nu}+\alpha K_{\mu\nu}=0=T_{\mu\nu},
\label{stealtheqs}
\end{equation}
where both side (the gravity and the matter parts) vanishes identically. Indeed, for
\begin{eqnarray}
\label{xi3}
\xi=\xi^{\tiny{\mbox{stealth}}}_{D}:=\frac{D-1}{4(D+1)},
\end{eqnarray}
and hence for a  potential (\ref{potstealth})
\begin{eqnarray}
\label{upot3}
U(\Phi)=\frac{(D-1)^2(D-3)}{32(D+1)}\Phi^2,
\end{eqnarray}
a solution of the stealth equations (\ref{stealtheqs})  is given by
\begin{eqnarray}
&&ds^{2}=-\left(r^2-\frac{M}{r^{\frac{D-5}{2}}}\right)dt^{2}+\frac{dr^2}{\left(r^2-\frac{M}{r^{\frac{D-5}{2}}}\right)}+{r^{2}}d\vec{x}_{D-2}^2\nonumber\\
\nonumber\\
&& \Phi(r)=A\,r^{\frac{1-D}{4}},
\label{solstealthk2}
\end{eqnarray}
where $A$ is an arbitrary constant. First of all, and contrary to the previous solution, in the zero  mass limit $M=0$, the scalar field does not vanish, and the solution reduces to a stealth configuration on pure AdS spacetime \cite{ABMTZ}. We also note that the metric solution (\ref{solstealthk2}) corresponds to the planar version ($\gamma=0$) of the solution obtained in Ref. \cite{Crisostomo:2000bb, Aros:2000ij} for the
gravity action given by (\ref{actionBHscan}) \footnote{Note that solutions of the Einstein-Gauss-Bonnet equations with arbitrary coupling constants were first obtained in the case $\Lambda=0$ with a spherical topology in \cite{Boulware:1985wk} and in \cite{Cai:2001dz} for arbitrary topology.}. In other words, we have derived a black hole stealth configuration for a
self-interacting scalar field with potential (\ref{upot3}) nonminimally coupled with parameter $\xi$ given by (\ref{xi3}) on a spacetime geometry that is the
planar solution of the particular Einstein-Gauss-Bonnet gravity action,  \cite{Crisostomo:2000bb, Aros:2000ij}. The occurrence of such solution can be explained easily. In fact, it is not difficult
to establish that a self-interacting scalar field with potential (\ref{potstealth}) given by
\begin{eqnarray}
\label{phiste}
\Phi(r)=A\,r^{\frac{2\xi}{4\xi-1}},
\end{eqnarray}
has a vanishing energy-momentum tensor $T_{\mu\nu}=0$ on the following $\xi-$dependent geometry
\begin{eqnarray} \label{geostel}
&&ds^{2}=-F(r)dt^{2}+\frac{dr^2}{F(r)}+r^2d\vec{x}_{D-2}^2,\nonumber\\
&&F(r)=\Big(r^2-\frac{M}{r^{\frac{4(D-2)\xi-(D-3)}{4\xi-1}}}\Big)
\end{eqnarray}
On the other side, as shown in \cite{Crisostomo:2000bb, Aros:2000ij}, the metric function $F(r)=r^2-\frac{M}{r^{\frac{D-5}{2}}}$ satisfies the gravity equation
$G_{\mu\nu}+\Lambda g_{\mu\nu}+\alpha K_{\mu\nu}=0$ for planar base manifold at the point (\ref{L2}). Hence, requiring that both side of the stealth equations (\ref{stealtheqs}) must vanish,
this will fix the value of the parameter $\xi$ to be (\ref{xi3}). As a final remark concerning the black hole stealth, it is interesting to note that the stealth metric (\ref{geostel}) will correspond to the
Schwarzschild-AdS-Tangherlini metric with planar base manifold only for $\xi=0$, but this case is of little interest since for a vanishing coupling parameter,
the scalar field becomes constant (\ref{phiste}).

Hence, we have obtained two classes of solutions for the Einstein-Gauss-Bonnet equations at the point (\ref{L2}) with a matter source
composed by a self-interacting scalar field nonminimally coupled. These solutions have been derived for particular values of the nonminimal coupling
parameter $\xi=\xi^{\tiny{\mbox{b.h}}}_{D}$ or $\xi=\xi^{\tiny{\mbox{stealth}}}_{D}$, and in both cases $\xi<1/4$. It is also intriguing to note that these coupling as well as the metric function solutions are related through the dimension as
$$
\xi^{\tiny{\mbox{b.h}}}_{D+1}=\xi^{\tiny{\mbox{stealth}}}_{D},\qquad F^{\tiny{\mbox{b.h}}}_{D+1}(r)=F^{\tiny{\mbox{stealth}}}_{D}(r).
$$

\section{Turning on the nonminimal parameter with exact $p-$forms}
For both class of solutions derived previously, the value of the nonminimal coupling parameter is unique and fixed in term of the dimension $D$, see (\ref{xi2}) and  (\ref{xi3}). This  feature is not a novelty and also occurs in Einstein gravity (with eventually a cosmological constant) with a scalar
field nonminimally coupled to gravity. Indeed, in this case, the only known black hole solutions are those obtained in four dimensions for the conformal coupling parameter $\xi=1/6$ and whose horizon
topology is either spherical or hyperbolic, see \cite{Bekenstein:1974sf, Bocharova:1970, Martinez:2002ru, Martinez:2005di, Anabalon:2012tu}. Recently,
it has been shown that the inclusion of multiple exact $p-$forms homogenously distributed permits the construction of black holes with planar horizon \cite{Bardoux:2012aw, Bardoux:2012tr}
without any restrictions on the dimension or on the value of the nonminimal parameter \cite{Caldarelli:2013gqa}. Indeed, in one side, an appropriate coupling between the
scalar field and the exact $p-$forms permits to relax the condition on the nonminimal parameter as well as the dimension. On the other side, the $p-$forms being homogenously distributed
imposes the horizon topology to be planar. Since, our working hypothesis is concerned with black hole solutions with planar base manifold, we propose to introduce appropriately some exact $p-$forms to obtain topological black hole solutions with arbitrary nonminimal coupling parameter.  In order to achieve this task, we will consider the following action in arbitrary $D$ dimension
\begin{eqnarray}
S-\int\! d^Dx\,\sqrt{-g}
\left[\frac{\epsilon(\Phi)}{2(D-1)!}\sum_{i=1}^{D-2}
{\cal H}^{(i)}_{\alpha_1\cdots \alpha_{D-1}}{\cal H}^{(i)\alpha_1\cdots \alpha_{D-1}}\right].\nonumber
\end{eqnarray}
Here $S$ denotes our original action (\ref{action1}) with the coupling constants choosing as (\ref{L2}) to which we have added  $(D-2)$ fields which are exact $(D-1)$-forms ${\cal H}^{(i)}$. The
coupling function between the scalar field and the $(D-1)$-forms denoted by $\epsilon(\Phi)$ depends on the scalar field $\Phi$ as
\begin{eqnarray}
\label{ep}
\epsilon(\Phi)=\sigma\,\Phi^{\frac{1-8\xi}{\xi}}
\end{eqnarray}
where $\sigma$ is a coupling constant. The field equations obtained by varying the action with the different dynamical fields $g_{\mu\nu}, \Phi$ and ${\cal H}^{(i)}$ read
\begin{widetext}
\begin{subequations}
\label{eqsmotionkpf}
\begin{eqnarray}
&&G_{\mu\nu}+\Lambda g_{\mu\nu}+\alpha K_{\mu\nu}=T_{\mu\nu}+T_{\mu\nu}^{\tiny{extra}},\qquad\qquad \nabla_{\alpha}\left(\epsilon {\cal H}^{(i)\alpha\alpha_1\cdots \alpha_{D-2}}\right)=0,\\
&&\Box \Phi = \xi R \Phi+\frac{d U}{d\Phi}+\frac{1}{2}\frac{d\epsilon}{d \Phi}\left[\sum_{i=1}^{D-2}\frac{1}{(D-1)!}{\cal H}^{(i)}_{\alpha_1\cdots \alpha_{D-1}}
{\cal H}^{(i)\alpha_1\cdots \alpha_{D-1}}\right]=0,
\end{eqnarray}
\end{subequations}
\end{widetext}
where the extra piece in the energy-momentum tensor reads
\begin{eqnarray}
T_{\mu\nu}^{\tiny{extra}}=&&\epsilon\sum_{i=1}^{D-2}\Big[\frac{1}{(D-2)!}
{\cal H}^{(i)}_{\mu\alpha_2\cdots \alpha_{D-1}}
{\cal H}_{\nu}^{(i)\alpha_2\cdots \alpha_{D-1}}\nonumber\\
&&-\frac{g_{\mu\nu}}{2(D-1)!}{\cal H}^{(i)}_{\alpha_1\cdots \alpha_{D-1}}
{\cal H}^{(i)\alpha_1\cdots \alpha_{D-1}}\Big].
\end{eqnarray}
Looking for a purely electrically homogenous Ansatz for the $(D-1)-$forms as
\begin{eqnarray*}
&&{\cal H}^{(i)}_{trx_1\cdots x_{i-1}x_{i+1}\cdots x_{D-2}}(r) dt dr\cdots dx^{i-1}dx^{i+1}\cdots dx^{D-2},
\end{eqnarray*}
where wedge product is understood. A solution of the field equations (\ref{eqsmotionkpf}) with a
purely electric Ansatz is given
\begin{eqnarray}
\label{solp}
ds^{2}=&-\left(r^2-\frac{M}{r^{\frac{2(6\xi-1)}{1-4\xi}}}\right)dt^{2}+
\frac{dr^2}{\left(r^2-\frac{M}{r^{\frac{2(6\xi-1)}{1-4\xi}}}\right)}+{r^{2}}d\vec{x}_{D-2}^2,\nonumber\\
\nonumber\\
&\Phi(r)=\sqrt{\frac{M(8\xi-1)(D-2)}{2\xi\left[2\xi(3D-4)-(D-2)\right]}}r^{\frac{2\xi}{4\xi-1}},\\
\nonumber\\
&{\cal H}^{(i)}_{trx_1\cdots x_{i-1}x_{i+1}\cdots x_{D-2}}=\frac{p}{\epsilon(\Phi)}\,r^{D-4},\nonumber
\end{eqnarray}
where the constant $p$ is given by
\begin{eqnarray}
&&p=\frac{4 M^{\frac{1-4\xi}{4\xi}}}{(4\xi-1)}\Big[\frac{2\xi}{(8\xi-1)(D-2)}\Big]^{\frac{8\xi-1}{4\xi}}\times\nonumber\\
&&\Big[2\xi(3D-4)-(D-2)\Big]^{\frac{6\xi-1}{4\xi}}\nonumber\times\\
&&\sqrt{-\sigma D(D+1)\xi(\xi-\xi^{\tiny{\mbox{b.h}}}_D)(\xi-\xi^{\tiny{\mbox{stealth}}}_D)},\nonumber
\end{eqnarray}
and where the constants $\xi^{\tiny{\mbox{b.h}}}_D$ and $\xi^{\tiny{\mbox{stealth}}}_D$ are the particular values given by (\ref{xi2}-\ref{xi3}).

As it should be, for $\xi=\xi^{\tiny{\mbox{b.h}}}_D$ or $\xi=\xi^{\tiny{\mbox{stealth}}}_D$, the constant $p=0$, and the resulting solutions
are those derived previously with source only given by a scalar field. Another interesting value is $\xi=1/8$ since in this case the scalar field vanishes and the coupling function $\epsilon$ becomes constant (\ref{ep}). As a consequence for $\xi=1/8$, we end with a pure axionic solution (that is a solution without scalar field)
$$
G_{\mu\nu}+\Lambda g_{\mu\nu}+\alpha K_{\mu\nu}=T_{\mu\nu}^{\tiny{extra}},
$$
given by
\begin{eqnarray}
&&ds^{2}=-\left(r^2-Mr\right)dt^2+
\frac{dr^2}{\left(r^2-Mr\right)}+{r^{2}}d\vec{x}_{D-2}^2,\nonumber\\
\nonumber\\
&&{\cal H}^{(i)}_{trx_1\cdots x_{i-1}x_{i+1}\cdots x_{D-2}}=-\sqrt{\frac{D-3}{2\sigma}}M\,r^{D-4}.\nonumber
\end{eqnarray}
Note that this last solution is obtained from the generic solution (\ref{solp}) taking the well-defined limit $\xi\to 1/8$.

Finally, it is interesting to mention that solutions of the field equations (\ref{eqsmotionkpf}) can be obtained without imposing the
values of the cosmological and Gauss-Bonnet coupling constants as given by (\ref{L2}) but rather considering the following relation between them
\begin{eqnarray}
\label{relgL}
\alpha=\frac{2\Lambda+(D-1)(D-2)}{(D-1)(D-2)(D-3)(D-4)}.
\end{eqnarray}
Of course, the restrictions (\ref{L2}) are a particular case of this last constraint. In fact, the relation (\ref{relgL}) is obtained by requiring
that the pure AdS metric solves the gravity equations without source. For a value of the parameter $\xi$ given by the conformal one in $(D+1)$ dimension, $\xi=\xi_{D+1}$, which in turn implies that the potential vanishes (\ref{potstealth}), a solution of the field equations (\ref{eqsmotionkpf}) can be obtained for a coupling $\epsilon$ given by
$$
\epsilon(\Phi)=\frac{\sigma}{\Phi^{\frac{4(D-2)}{D-1}}}.
$$
In this case, the metric function is the Schwarzschild-AdS-Tangherlini spacetime
\begin{eqnarray}
\label{solpxid+1}
&ds^{2}=-\left(r^2-\frac{M}{r^{D-3}}\right)dt^{2}+
\frac{dr^2}{\left(r^2-\frac{M}{r^{D-3}}\right)}+{r^{2}}d\vec{x}_{D-2}^2,\nonumber\\
\nonumber\\
&\Phi(r)=\sqrt{\frac{8MD(D-2)\Big(2\Lambda+(D-1)(D-2)\Big)}{(D^2-3D+4)(D-1)^2}}r^{\frac{1-D}{2}},\\
\nonumber\\
&{\cal H}^{(i)}_{trx_1\cdots x_{i-1}x_{i+1}\cdots x_{D-2}}=\frac{p}{\epsilon(\Phi)}\,r^{D-4},\nonumber
\end{eqnarray}
where the constant $p$ is given by
\begin{eqnarray*}
p=&&\frac{\sqrt{-2M\sigma}}{4(D-2)}\,\left(D-1\right)^{\frac{5D-9}{2(D-1)}}\times\\
&&\left(\frac{(D^2-3D+4)}{8MD(D-2)\Big(2\Lambda+(D-1)(D-2)\Big)}\right)^{\frac{D-3}{2(D-1)}}.
\end{eqnarray*}
We would like to stress that this solution is valid even in the vanishing cosmological constant but on the other hand, the GR limit
$\Lambda=-(D-1)(D-2)/2$ (\ref{relgL}) is not well-defined because the constant $p$ will blow up. This clearly emphasizes the importance of the higher-order
curvature terms present in the Einstein-Hilbert-Gauss-Bonnet Lagrangian.

\section{Conclusions and Further Works}
Here, the gravity action we have considered is a particular combination of the Einstein-Hilbert action
with a negative cosmological constant together with the Gauss-Bonnet density. In five dimensions, this action
corresponds to a Chern-Simons action which is a particular case of the so-called Lovelock action. This latter can be
viewed as a generalization of the Einstein gravity in arbitrary
dimension yielding at most to second-order field equations for the metric. The resulting theory is
described by  a $D-$form
constructed with the vielbein $e^a$, the spin connection
$\omega^{ab}$, and their exterior derivatives without using the
Hodge dual. The Lovelock action is a polynomial of degree
$[D/2]$ (where $[x]$ denotes the integer part of $x$) in the curvature two-form, $R^{ab} = d\,\omega^{ab} +
\omega^{a}_{\;c} \wedge \omega^{cb}$ as
\begin{eqnarray}
&&\int \sum_{p=0}^{[D/2]}\alpha_p~ L^{(p)},\\
&& L^{(p)}=\epsilon_{a_1\cdots
a_d} R^{a_1a_2}\cdots R^{a_{2p-1}a_{2p}}e^{a_{2p+1}}\cdots  e^{a_d},\nonumber
\end{eqnarray}
where the $\alpha_p$ are arbitrary dimensionful coupling constants
and where wedge products between forms are understood. Here $L^{(0)}$ and $L^{(1)}$ are respectively the
well-known cosmological term and the Einstein-Hilbert Lagrangian. As shown in Ref. \cite{Crisostomo:2000bb},
requiring the Lovelock  action to have a unique
cosmological constant, fixes the $\alpha_p$  yielding to a series of actions indexed by an integer $k$ given by
\begin{eqnarray}
I_k=-\int \sum_{p=0}^{k}\frac{C^k_p}{(D-2p)}~ L^{(p)},\quad 1\leq k\leq \Big[\frac{D-1}{2}\Big],
\label{Ik}
\end{eqnarray}
where $C^k_p$ corresponds to the combinatorial factors. The gravity action we have considered here (\ref{actionBHscan}) is nothing but
the action $I_2$ given by the expression (\ref{Ik}). We believe that the class of solutions derived here for $I_2$ can be generalized
for an arbitrary gravity action $I_k$ with $k\geq 2$, \cite{BH}. For example, the stealth black hole solution obtained for $I_2$ can easily be extended for
an arbitrary
action $I_k$ by adjusting the value of the nonminimal coupling parameter appearing in the stealth metric solution (\ref{geostel}) in order for the resulting metric to match with the pure gravity solution of $I_k$ \cite{Aros:2000ij, Crisostomo:2000bb}. Moreover, at our knowledge, there does not exist black hole solutions with planar base manifold
for standard General Relativity (with or without a cosmological constant) with a source given by a scalar field nonminimally coupled \footnote{ There exist solutions adding exact $p-$forms coupled with the scalar field, see \cite{Bardoux:2012tr}, \cite{Caldarelli:2013gqa}.}. This
reinforces our conviction that the existence of the solutions derived for  a gravity action given by $I_2$ are strongly inherent of the presence of the higher-order curvature terms, \cite{BH}.

We now turn to the choice of the mass term potential (\ref{potstealth}) in our starting action. First of all, we may note that this kind of
potential has been considered in the dual description of  unconventional superconductor \cite{Hartnoll:2009sz,Horowitz:2010gk,Hartnoll:2008kx,Herzog:2009xv}.
Moreover, in our case, because of the presence of the
nonminimal coupling term $\xi R\Phi^2$ in the action, in case of constant scalar curvature solutions $R=-D(D-1)$, one may define an effective square mass
$$
m_{{\tiny\mbox{eff}}}^2=-\xi D(D-1)\left[1-\frac{16}{(1-4\xi)^{2}}\,(\xi-\xi_{D})(\xi-\xi_{D+1})\right].
$$
In the solutions obtained here, the only ones of constant scalar curvature (apart from the trivial situation of taking $M=0$) are those
with the $p-$form fields (\ref{solp}) for $\xi=D/(4D+4)$ and $\xi=(D-1)/(4D)$. It is interesting to note that the square effective mass $m_{{\tiny\mbox{eff}}}^2$ precisely saturates the Breitenlohner-Freedman bound for $\xi=(D-1)/(4D)$. To close the chapter concerning the potential, we would like to stress that
is a particular case of potentials allowing the existence of  self-interacting scalar field $\Phi$ nonminimally coupled with vanishing stress tensor
\begin{eqnarray*}
T_{\mu \nu}:=&\partial_{\mu}\Phi\partial_{\nu} \Phi-g_{\mu
\nu}\Big(\frac{1}{2}\,\partial_{\sigma}\Phi\partial^{\sigma}\Phi+U(\Phi)\Big)\\
&+\xi\left(g_{\mu \nu}\Box-\nabla_{\mu}\nabla_{\nu}+G_{\mu\nu}\right)\Phi^{2}=0
\end{eqnarray*}
on the AdS background
\begin{eqnarray*}
ds^2=-r^2dt^2+\frac{dr^2}{r^2}+r^2d\vec{x}_{D-2}^2.
\end{eqnarray*}
Indeed, as shown in \cite{ABMTZ}, the stealth solution is given by the following configuration
\begin{subequations}
\label{scf1}
\begin{eqnarray}
\label{potg}
U(\Phi)=&\frac{\xi}{{(1-4\xi)^2}}\,\Big[2\,\xi\,b^{2}\Phi^{\frac{1-2\xi}{\xi}}-8\,(D-1)\left(\xi-\xi_{D}\right)\times \nonumber\\
        &\left(2\,\xi\,b\,\Phi^{\frac{1}{2\xi}}-D \left(\xi-\xi_{D+1}\right)\,\Phi^{2} \right)\Big],\\
\label{sf2}
&\Phi(r)=(Ar+b)^{\frac{2\xi}{4\xi-1}}
\end{eqnarray}
\end{subequations}
Note that this kind of potential also appears when looking for AdS wave solutions for a scalar filed nonminimally coupled \cite{AyonBeato:2005qq}. The scalar field solutions obtained in this paper as well as the potential considered correspond to the $b=0$ limit of this stealth
configuration (\ref{scf1}). It has been shown recently that, in the context of standard General Relativity, self-interacting scalar fields nonminimally coupled with potential given by (\ref{potg}) and extra axionic fields admit black hole solutions, \cite{Caldarelli:2013gqa}. The derivation of these solutions was precisely operated from the stealth configuration (\ref{scf1}) through a Kerr-Schild transformation, \cite{Caldarelli:2013gqa}. We then
believe that there may exist more general black hole solutions than those derived here for this more general
class of potentials (\ref{potg}). To conclude with the stealth origin, we would like to stress that the horizon topology of our solutions
is planar and, their extension to spherical or hyperboloid black hole solutions in a standard fashion was not possible. This may be
related to the fact that static stealth scalar field on the AdS background requires the base space to be flat \cite{ABMTZ} for dimensions $D\geq 4$.

An interesting task to realize will be the study of the thermodynamics properties of the solutions derived here for $I_2$ as well as those for
general $I_k$, and to compare them with the pure gravity solutions \cite{Aros:2000ij}.

In Ref. \cite{Oliva:2011np}, the authors constructed conformal coupling to arbitrary higher-order Euler densities. It will be interesting to see whether
such matter source can accommodate the kind of solutions derived here.

As a final remark, we note that the coupling with the Maxwell electromagnetic field is an open problem even in the black hole stealth case. Indeed, even if the pure gravity solution with Maxwell source
$$
G_{\mu\nu}+\Lambda g_{\mu\nu}+\alpha K_{\mu\nu}=T_{\mu\nu}^{\tiny{\mbox{Maxwell}}}
$$
is known \cite{Crisostomo:2000bb, Aros:2000ij}, it seems that this solution can not be promoted to a black hole stealth configuration with a nonminimal scalar
field as it was possible in the neutral case.

\begin{acknowledgments}
We thank Julio Oliva and Sourya Ray for useful discussions. MB is supported by BECA DOCTORAL CONICYT 21120271.
MH was partially supported by grant 1130423 from FONDECYT, by grant ACT 56 from CONICYT and from
CONICYT, Departamento de Relaciones Internacionales ``Programa Regional MATHAMSUD 13 MATH-05''.
\end{acknowledgments}


\end{document}